\documentclass[12pt,a4paper]{article}
\usepackage[cp1250]{inputenc}
\usepackage[english]{babel}
\usepackage[dvips]{graphicx}
\usepackage{amssymb}
\usepackage{amsmath}
\usepackage{amsfonts}

\def\bd{ \textrm{d}}
\def\eps{\epsilon}
\def\dz{\frac{\partial}{\partial z}}
\def\dy{\frac{\partial}{\partial y}}
\def\dyz{\frac{\partial^2}{\partial y \partial z}}
\def\dzz{\frac{\partial^2}{\partial z^2}}
\def\dyy{\frac{\partial^2}{\partial y^2}}
\def\aaa{A(v,t,s)}

\def\bba{A(v-1,t-1,s)}
\def\bab{A(v-1,t,s-1)}

\def\bac{A(v-1,t,s-2)}
\def\aac{A(v,t,s-2)}
\def\cha{\chi_{\{s \geq 1\}}}
\def\chb{\chi_{\{s \geq 2\}}}

\begin{document}

\title{Application of Generating Functions and Partial Differential Equations in Coding Theory}
\author{Milan Bradonji\'c
\\
{\small Electronic address: milan@ee.ucla.edu}\\
Department of Electrical Engineering\\
University of California, Los Angeles, CA 90095, USA
}

\date{June 2003\footnote{
This work was done in 2003, while the author was a
pre-doctoral student at EPFL, Swiss Federal Institute of
Technology, Switzerland}
}

\maketitle

\paragraph{Keywords:}{Cycle Poisson Case, LDPC coding, generating functions, partial
differential equations,Hadamarad Multiplication's Theorem.}


\section*{Abstract}

In this work we have considered formal power series and partial
differential equations, and their relationship with Coding Theory.
We have obtained the nature of solutions for the partial
differential equations for Cycle Poisson Case. The coefficients
for this case have been simulated, and the high tendency of growth
is shown. In the light of Complex Analysis, the Hadamard
Multiplication's Theorem is presented as a new approach to divide
the power sums relating to the error probability, each part of
which can be analyzed later.

\section{Generating Functions and Operations on them}

Let $A(z)$ and $B(z)$ be the two generating functions, in fact two
formal power sums [2], [3], whose coefficients belong to some
field, $a_n, b_n \in \mathbb{F}$:
\begin{equation}
A(z)=\sum_{n \geq 0} a_n z^n,
\end{equation}
and
\begin{equation}
B(z)=\sum_{n \geq 0} b_n z^n,
\end{equation}

The infinite series $\sum_{n \geq 0} a_n z^n$, absolutely
converges if and only if  there is a bounding constant $M$, such
that the finite sums $\sum_{0 \leq n \leq N} |a_n z^n|$ never
exceed $M$, for every $N \in \mathbf{N}$. It directly follows,
that, if $\sum_{n \geq 0} a_n z^n$ converges for some value
$z=z_0$, it also converges for all $z$ with $z<z_0$.  Even if the
series do not converge, the next operations we perform on
generating functions can be justified rigorously as an operation
on formal power series. Then we have the following are satisfied:

-Right shift
\begin{equation}
zA(z)=\sum_{n \geq 1} a_{n-1} z^n,
\end{equation}

-Left shift
\begin{equation}
\frac{A(z)-a_0}{z}=\sum_{n \geq 0} a_{n+1} z^n,
\end{equation}

-Differentiation
\begin{equation}
A'(z)=\sum_{n \geq 0} (n+1)a_{n+1} z^n,
\end{equation}

-Integration
\begin{equation}
\int_{0}^{z}A(t) \bd t = \sum_{n \geq 1} \frac{a_{n-1}}{n}z^n,
\end{equation}

-Scaling
\begin{equation}
A(\lambda z)=\sum_{n \geq 0} \lambda ^n a_n z^n,
\end{equation}

-Addition
\begin{equation}
A(z)+B(z)=\sum_{n \geq 0} (a_n+b_n) z^n,
\end{equation}

-Difference
\begin{equation}
(1-z)A(z)=a_0+\sum_{n \geq 1} (a_n-a_{n-1}) z^n,
\end{equation}

-Convolution
\begin{equation}
A(z)B(z)=\sum_{n \geq 0} \big( \sum_{0 \geq k \geq n} a_k b_{n-k}
\big) z^n,
\end{equation}

-Partial sum
\begin{equation}
\frac{A(z)}{1-z}=\sum_{n \geq 0} \big( \sum_{0 \geq k \geq n} a_k
\big) z^n.
\end{equation}

\section{Introduction}
\subsection{The Cycle Poisson Case}

In the next chapter we will consider the Poisson ensemble [1],
i.e., the code where the following conditions are satisfied:

-the length of the code is $n$,

-the rate is $r$,

-all variable nodes have degree two.

Every edge in the above defined code, is chosen i.i.d. with
uniform probability for all $(1-r)n$ check nodes. In the following
$m=(1-r)n$.

For the fixed number of variable nodes $v$, we have the following
number of constellations:

\begin{displaymath}
N(v)=\left \{ \begin{array}{ll}
((1-r)n)^{2v}, & 0 \leq v \leq n, \\
0, & \textrm{otherwise}.
\end{array} \right.
\end{displaymath}

The number of such constellations which are stopping sets on $t$
check nodes is equal to the following:

\begin{equation}
S(v,t):= {(1-r)n \choose t}(2v)! \textrm{coef}\{ (e^x-1-x)^t,
x^{2v} \}.
\end{equation}

If we have $v$ fixed nodes, having $t$ check nodes of degree all
least one, and there is no empty stopping-sets, the number of
constellations is equal to  $v!2^v\aaa$.

Then, we can derive the error probability
$\mathbb{E}[P^{IT}_B(G,\eps)]$:
\begin{eqnarray*}
\mathbb{E}[P^{IT}_B(G,\eps)] & = & \sum_{v}{n \choose v}(1-\eps)^{n-v} \frac{v!2^v\sum_{t,s}\aaa}{N(v)} \\
& = &  (1-\eps)^n \sum_{v}{n \choose v} v! \Big( \frac{2 \eps} {(1-\eps)((1-r)n)^2} \Big) ^v \sum_{t,s} \aaa \\
& = &  (1-\eps)^n \mathcal{A}(x= \frac{2 \eps} {(1-\eps)((1-r)n)^2} , y=1, z=1), \\
\end{eqnarray*}
where
\begin{eqnarray*}
\mathcal{A}(x,y,z) & = & \sum_{v,t,s}\frac{x^v}{(n-v)!} y^t z^s,  \\
& = &  \sum^{n}_{v=1} \sum^{m}_{t=1} \sum^{m-t}_{s=0} \aaa
\frac{x^v}{(n-v)!} y^t z^s,
\end{eqnarray*}
is the generating function with coefficients $\aaa / (n-v)!$.


For indices, $1 \leq v \leq n, 1 \leq t \leq m, 0 \leq s \leq
m-t$, the following is satisfied:
\begin{eqnarray*}
\aaa s & = & \bac(m-t-s+2)(m-t-s+1) \chb  \\
       & + & \bab (m-t-s+1)t  \cha \\
       & + & \bba (m-t-s+1)s  \cha.
\end{eqnarray*}

For easy  of notations let $k=m+1$. Now, we start with the
recurrence equation, but considering the generating function:
\begin{eqnarray}
G(x,y,z)=\sum_{v,t,s} \aaa x^v y^t z^s,
\end{eqnarray}
except $\mathcal{A}(x,y,z)$.

\subsection{Derivation of PDE}
Using rules derived in the previous section, which relate to the
operations on the generating functions, we get:

\begin{eqnarray*}
k(k+1) \aac & \equiv & k(k+1)z^2, \\
2kt    \aac & \equiv & 2kyz^2 \dy, \\
2ks    \aac & \equiv & 2kz^2(2+z \dz), \\
t(t-1) \aac & \equiv & y^2z^2 \dyy,\\
s(s-1) \aac & \equiv & z^2(2+4z \dz +z^2 \dzz), \\
2ts    \aac & \equiv & 2yz^2(2 \dy + z \dyz).
\end{eqnarray*}

Putting things together, we get:
\begin{eqnarray*}
z \dz G & = & x \Big\{ k(k+1)z^2 -2kyz^2 \dy - 2kz^2(2+z \dz) \\
        & + & y^2z^2 \dyy +z^2(2+4z \dz +z^2 \dzz ) + 2yz^2 (2 \dy +z \dyz) \\
        & + & myz \dy -zy^2 \dyy  -yz(\dy+z \dyz) + myz\dz \\
    & - & yz^2\dzz -yz(\dz+y\dyz) \Big\} G.
\end{eqnarray*}

On the other hand,
\begin{eqnarray*}
z \dz G & = & x \Big\{ k(k+1)z^2 -4kz^2 +2z^2\\
        & + & \{ -2kz^2 + 4yz^2 +myz -yz \} \dy \\
        & + & \{ -2kz^2 +4z^3 +myz -yz\} \dz \\
    & + & \{ y^2z^2-zy^2\} \dyy \\
    & + & \{ z^4 -yz^2 \} \dzz \\
    & + & \{ 2yz^3-yz^2-y^2z\} \dyz \Big\} G.
\end{eqnarray*}

Our partial differential equation has just been derived:
\begin{eqnarray*}
z \dz G & = & xz \Big\{ (k^2-3k+2)z + y(2-k)(2z-1) \dy  \\
    & + & (2-k)(2z^2-y) \dz + y^2(y-1) \dyy \\
    & + & z(z^2-y) \dzz +  y(2z^2-y-z) \dyz \Big\} G,
\end{eqnarray*}
where $\dy, \dz, \dyz, \dyy, \dzz$ are the partial derivatives,
which are applied on the function $G$.

\subsection{Conclusions and Improvement}
What is the capital gain derived from the previous analysis? The
crucial thing is that, considering the generating function
$G(x,y,z)$ over coefficients $A(v,t,s)$, instead of the generating
functions $\mathcal{A}(x,y,z)$, we have got the second order
partial differential equation, where $x$ is a parameter.

\section{Partial Equation and Determination of Solutions}
For $z=0$ we have identity. If $z$ is different than zero, we have
to consider:

a) $x=0$, that will imply $\frac{\partial G}{\partial z}=0$, which
is the boundary condition for the plane  $(x,y,z) \in \{
(0,\lambda,\mu) | \lambda \in \mathbf{R}, \mu \neq 0 \}$;

b) Let us suppose $x \neq 0$ and $z \neq 0$, then we get the
following partial differential equation:

\begin{equation}
\frac{1}{x}\frac{\partial G}{ \partial z} = AG_{yy}+ 2BG_{yz} + C
G_{zz} +D G_y + E'G_z + F.
\end{equation}

\begin{equation}
AG_{yy}+ 2BG_{yz} + C G_{zz} +D G_y + (E' -\frac{1}{x})G_z + F =0,
\end{equation}
where polynomials $A, B, C, D, E', F$ are given respectively as:

\begin{eqnarray*}
A(y)    & = & y^2(y-1), \\
B(y,z)  & = & \frac{1}{2} y(2z^2-y-z), \\
C(y,z)  & = & z(z^2-y), \\
D(y,z)  & = & y(2-k)(2z-1), \\
E'(y,z) & = & (2-k)(2z^2-y), \\
F(z)    & = & (k^2-3k+2)z.
\end{eqnarray*}

Using term $B^2-AC$ we can determine the nature of the solution,
[6], [7]:

\begin{displaymath}
B^2-AC = \left \{ \begin{array}{ll}
>0, & \textrm{hyperbolic} \\
=0, & \textrm{parabolic} \\
<0, & \textrm{elliptic} \\
\end{array} \right.
\end{displaymath}

In that case, consider $(2B)^2-4AC$. We get the following:

\begin{eqnarray*}
(2B)^2-4AC & = & y^2(z^2-y-z)^2 - y^2(y-1)z(z^2-y) \\
       & = & y^2 ( 4z^4 -3z^3 +z^2 + y^2 -yz^3 - 3 yz^3 +yz ).
\end{eqnarray*}

In the aim of determining the nature of the solution, let $y =
\alpha z$. With this substitution, we are able to divide the $y-z$
plane in disjunct regions, where each region belongs to one type
of the solution. We have:

\begin{eqnarray*}
4(B^2-AC)  & = & y^2 ( 4z^4 -3z^3 +z^2 + y^2 -yz^3 - 3 yz^3 +yz ) \\
       & = & z^4 \alpha ^2 ( (4-\alpha)z^2 - 3(1+\alpha)z + 1 + \alpha + \alpha^2 ).
\end{eqnarray*}

Consider the function, $f(z)=(4-\alpha)z^2 - 3(1+\alpha)z + 1 +
\alpha + \alpha^2$, whose discriminant is:

\begin{eqnarray*}
D(\alpha) & = & 9(1+\alpha)^2 - 4(4-\alpha)(1+\alpha+\alpha^2), \\
      & = & (\alpha-1)(4\alpha^2+\alpha+7).
\end{eqnarray*}

Now, $f(z)=(4-\alpha)z^2 - 3(1+\alpha)z + 1 + \alpha + \alpha^2$,
also we have $(4\alpha^2+\alpha+7) \geq 0$, for $\alpha \in
\mathbf{R}$, and $4D_{PEQ}=4(B^2-AC)=\alpha^2 z^4 f(z)$.

Let $z_{1,2}$ be the roots of the equation $f(z)=0$, so we get the
following:

0) $ \alpha =0,$ we have a strictly parabolic case,
\begin{displaymath}
\left \{ \begin{array}{ll} \forall z \in \mathbf{R}, D_{PEQ}=0,  &
\textrm{parabolic}
\end{array} \right.
\end{displaymath}

1) $ 0 < \alpha < 1,$
\begin{displaymath}
\left \{ \begin{array}{ll} \forall z \in \mathbf{R}, D_{PEQ}>0,  &
\textrm{hyperbolic}
\end{array} \right.
\end{displaymath}

\begin{figure}[h]
\begin{center}
\includegraphics[width=5cm]{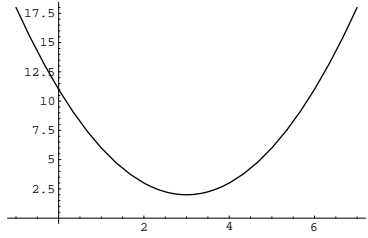}
\end{center}
\caption{\label{figur1q} Function $f(z)$}
\end{figure}

\newpage
2) $ \alpha = 1,$
\begin{displaymath}
\left \{ \begin{array}{ll}
z=3, D_{PEQ}=0, &      \textrm{parabolic} \\
z \neq 3, D_{PEQ}>0, & \textrm{hyperbolic} \\
\end{array} \right.
\end{displaymath}
\begin{figure}[h]
\begin{center}
\includegraphics[width=5cm]{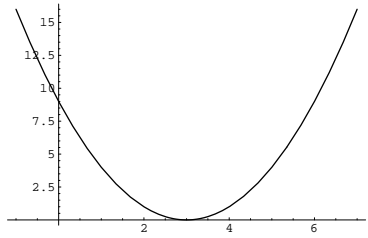}
\end{center}
\caption{\label{figur1b} Function $f(z)=(z-3)^2$}
\end{figure}

3) $ 1< \alpha < 4,$
\begin{displaymath}
\left \{ \begin{array}{ll}
z \in [z_1,z_2], D_{PEQ}<0, &         \textrm{elliptic} \\
z = z_1 \lor z= z_2, D_{PEQ}=0, &     \textrm{parabolic} \\
z \notin [z_1,z_2], D_{PEQ}>0, &      \textrm{hyperbolic} \\
\end{array} \right.
\end{displaymath}
\begin{figure}[h]
\begin{center}
\includegraphics[width=5cm]{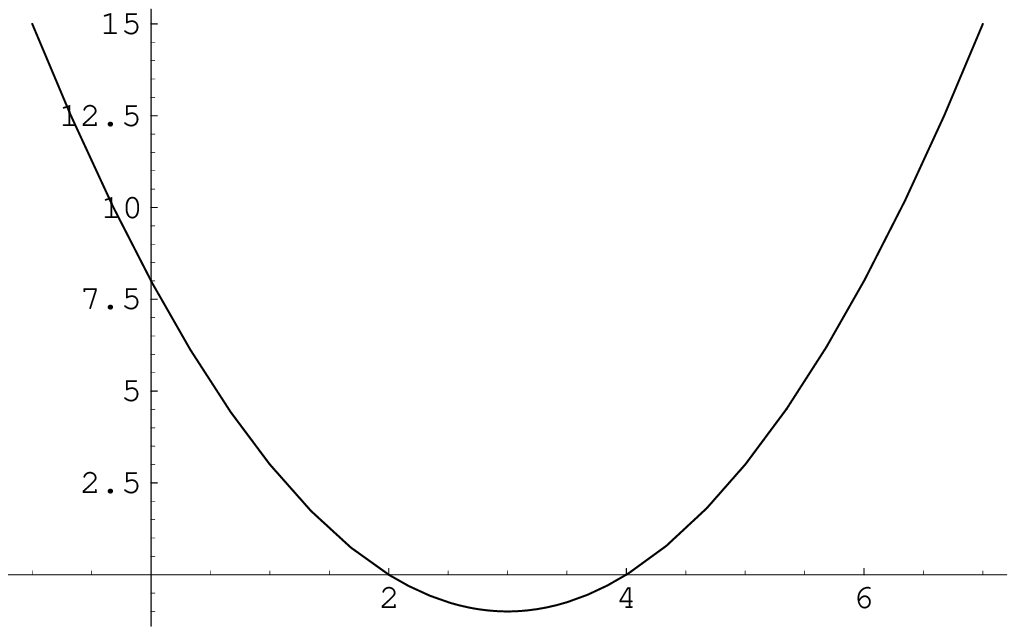}
\end{center}
\caption{\label{figur1p} Function $f(z)$, E.g. $z_1=2, \quad
z_2=4$}
\end{figure}

\newpage
4) $ \alpha =4,$
\begin{displaymath}
\left \{ \begin{array}{ll}
z > \frac{7}{5}, D_{PEQ}<0, &      \textrm{elliptic} \\
z = \frac{7}{5}, D_{PEQ}=0  &      \textrm{parabolic} \\
z < \frac{7}{5}, D_{PEQ}>0, &      \textrm{hyperbolic} \\
\end{array} \right.
\end{displaymath}
\begin{figure}[h]
\begin{center}
\includegraphics[width=5cm]{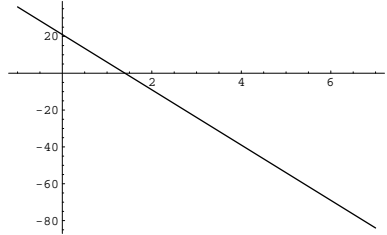}
\end{center}
\caption{\label{figur1b2} Function $f(z)=-15z+21$}
\end{figure}

5) $ \alpha >4,$
\begin{displaymath}
\left \{ \begin{array}{ll}
z \in [z_1,z_2], D_{PEQ}>0, &          \textrm{hyperbolic} \\
z = z_1 \lor z= z_2, D_{PEQ}=0, &      \textrm{parabolic} \\
z \notin [z_1,z_2], D_{PEQ}<0, &       \textrm{elliptic} \\
\end{array} \right.
\end{displaymath}
\begin{figure}[h]
\begin{center}
\includegraphics[width=5cm]{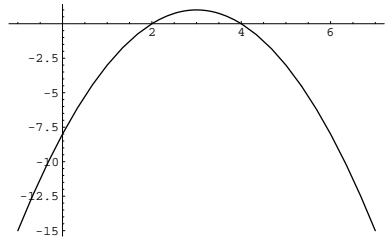}
\end{center}
\caption{\label{figur1r} Function $f(z)$, E.g. $z_1=2, \quad
z_2=4$}
\end{figure}

\newpage
Summary, for the nature of the our partial differential
equation, we are getting the following:

\begin{figure}[h]
\begin{center}
\includegraphics[width=12cm]{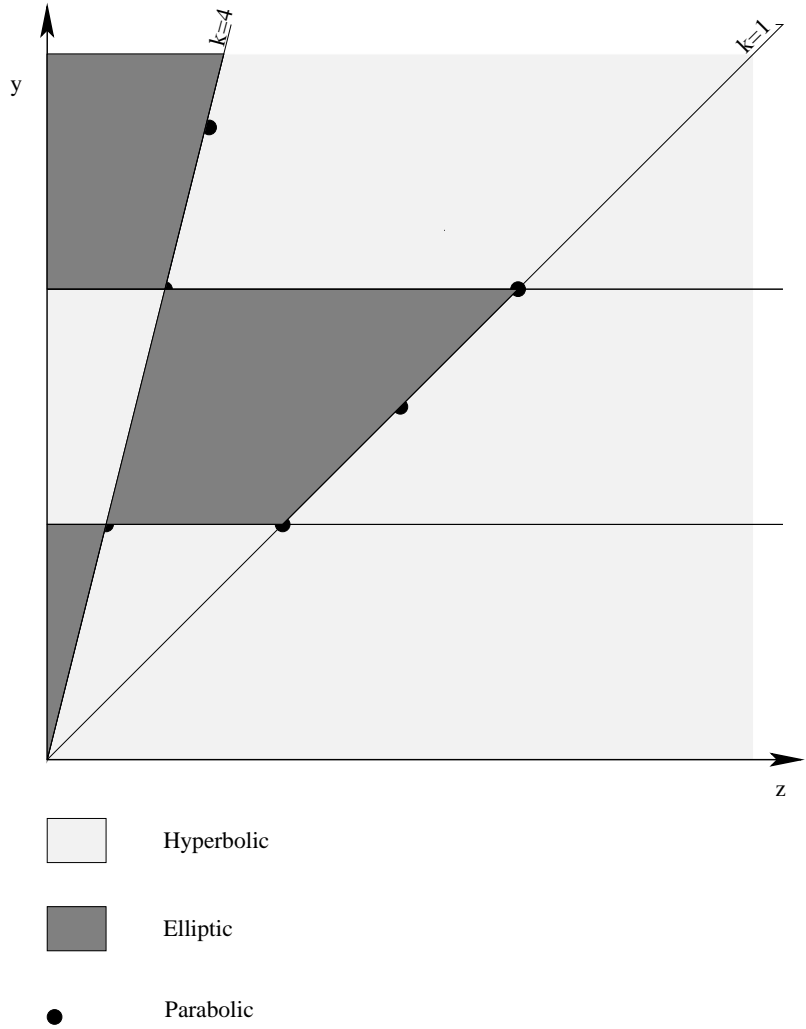}
\end{center}
\caption{\label{figur1u} The Nature of PDE }
\end{figure}

\section{Boundary Conditions}

For the boundary condition, the coefficients $A(v,t,s=0)$, satisfy
the following:
\begin{equation}
A(v,t,s=0) = {(1-r)n \choose t}(2v)! \textrm{coef}\{ (e^x-1-x)^t,
x^{2v} \}.
\end{equation}
Since $\frac{(2v)!}{2^v v!}=(2v-1)!!$, we have:
\begin{equation}
A(v,t,s=0) = {m \choose t}(2v-1)!! \textrm{coef}\{ (e^x-1-x)^t,
x^{2v} \},
\end{equation}
where $m=(1-r)n$.

In the aim of getting the exponents for the given coefficients, we
have simulated $\log \frac{ A(v,t,s=0)}{{m \choose t}}$ with
increasing $t$. It is done for $m=100$, and $t$ taking the values
$1,2,3,4,5,10,30,20,40,50$. It is important to notice, that given
coefficients do not depend on the length and on the rate of the
code individually, but on $m=(1-r)n$. The binomial coefficient ${m
\choose t}$ can be derived from the Stirling's formula:

\begin{equation}
n! \approx \sqrt{(2n+\frac{1}{3}) \pi } \Big( \frac{n}{e}\Big)^n,
\end{equation}
for $n$ sufficient large.

On the next page are given simulated values, where $g^{(t)}(v)$ is
the exponential factor:
\begin{equation}
g^{(t)}(v)=\log \frac{A(v,t,s=0)}{{m \choose t}},
\end{equation}
for the appropriate $t$.

In the Appendix, it will be shown the simulated exponential
factors $g^{(t)}(v)$, for different $t$'s.

\section{Another views on the problem}
\subsection{Complex Analysis}

In this section we will use another approach, based on Complex
Analysis. The following problem was considered by Hadamard, [4].
\\
\textbf{Theorem}(\textit{Hadamard's Multiplication Theorem})

Suppose that,

\begin{equation}
f(z)=\sum_{n=0}^{+\infty} a_n z^n,
\end{equation} is convergent for $|z|<R$, and

\begin{equation}
g(z)=\sum_{n=0}^{+\infty} b_n z^n,
\end{equation}  is convergent for $|z|<R'$
and that the singularities of $f(z)$ and $g(z)$ are known.

Consider the function:

\begin{equation}
F(z)=\sum_{n=0}^{+\infty} a_n b_n z^n.
\end{equation}

$F(z)$ is regular for $|z|<RR'$, and Hadamard's theorem depends on
the following representation of $F(z)$ as an integral:
\begin{equation}
F(z)=\frac{1}{2 \pi i } \int_c f(w)g \Big( \frac{z}{w} \Big)
\frac{\bd w}{w},
\end{equation}
where $C$ is a contour, including the origin, on which $|w|<R,
|z/w|<R'$.

To prove this, write
\begin{equation}
g \Big( \frac{z}{w} \Big)=\sum_{n=0}^{\infty} b_n \Big(\frac{z}{w}
\Big)^n,
\end{equation}
in the integral, and integrate term by term, as we may by uniform
convergence. We obtain:

\begin{eqnarray*}
F(z) & = & \frac{1}{2 \pi i } \int_c f(w)g(\frac{z}{w})\frac{\bd w}{w} \\
     & = & \sum_{n=0}^{+\infty}\frac{b_n z^n}{2\pi i} \int_c \frac{f(w)}{w^{n+1}}\bd w \\
     & = &\sum_{n=0}^{+\infty} a_n b_n z^n,
\end{eqnarray*}
the required result.

\subsection{Returning to Error Probability}

\begin{eqnarray*}
E_B & = & (1-\eps)^n \sum_{v}{n \choose v} v! \Big( \frac{2 \eps} {(1-\eps)((1-r)n)^2} \Big) ^v \sum_{t,s} \aaa \\
    & = & (1-\eps)^n \sum_{t,s} \underbrace {\sum_{v}{n \choose v} v!  \aaa \frac{1}{n^{2v}} \Big( \frac{2 \eps} {(1-\eps)((1-r))^2} \Big) ^v}_{S}.
\end{eqnarray*}

Let $x$ be $x=\frac{2 \eps} {(1-\eps)(1-r)^2}$, and consider the
sum:

\begin{equation}
S  =  \sum_{v}{n \choose v} v! \aaa \frac{1}{n^{2v}} x ^v.
\end{equation}

The following is known, [5], [8], [9]:

\begin{equation}
 \sum_{v} {n \choose v}\frac{x^v}{n^{2v}}=(1+\frac{x}{n^2})^n, \quad \textrm{for $|x|<1$},
\end{equation}
\begin{equation}
\sum_{v} {n \choose v}x^v = (1+x)^n ,  \quad \textrm{for $|x|<1$},
\end{equation}
\begin{equation}
\sum_{v} v!x^v, \quad \textrm{diverge for all $x$}.
\end{equation}

If we could bound the series:

\begin{equation}
 \sum_{v} v! \aaa x^v
\end{equation}

\begin{equation}
\sum_{v} \frac{{v! \aaa}}{n^{2v}}x^v,
\end{equation}

\begin{equation}
\sum_{v} \frac{{n \choose v} \aaa}{n^{2v}}x^v,
\end{equation}
then Hadamard's multiplications theorem could be applied to the
given series,  and some bound could be found. It is very important
to remark here, that the derived bound would be for  a specified
$t$ and $s$. It follows that only tight inequalities would satisfy
us, because we need to sum over all $t$ and $s$, in the aim of
getting $E_B$.

\newpage
\section{Conclusions}

In this work we have shown, how formal power series can be used
and how one can operate on them, even if the series do not
converge.

We have been analyzing the Cycle Poisson Case, looking for the
error probability given by the recurrence equation, whose
coefficients were dependent up to a depth two of indices.

Using appropriate generating functions, we have derived second
order partial differential equation. It is shown that $x$ took
part like a parameter, and the nature of the solution was
dependent only on variables $y$ and $z$, but not on $x$. The
nature of the solutions, i.e. how equation behaves is completely
determined.

To know how the first coefficients look like, we simulated them.
The high tendency of growing is obtained, even for small values of
$t$, the coefficients are going up to $10^{200}$.

Then we suggested totally new approach in the domain of Complex
Analysis. The problem is divided into parts, looking for two
appropriate series, whose scalar-multiplication form desired power
sum, and try to bound both of them.

It is obvious that for examples whose recurrence equations derive
some solvable partial differential equation, given method
succeeds, otherwise it is on us how to approach the solution.

\section{Appendix}
\subsection{Diagrams of Exponential Factors}

On the following two pages are shown the simulated exponential
factors $g^{(t)}(v)$, for different $t$'s, derived in the Section
4, \textit{Boundary Conditions}.

\begin{figure}[hp]
\begin{center}
\begin{minipage}[h]{5.0cm}
\vspace{5.0cm}
\begin{picture}(5.0,5.0)
\includegraphics[height=5.0cm,width=5.0cm]{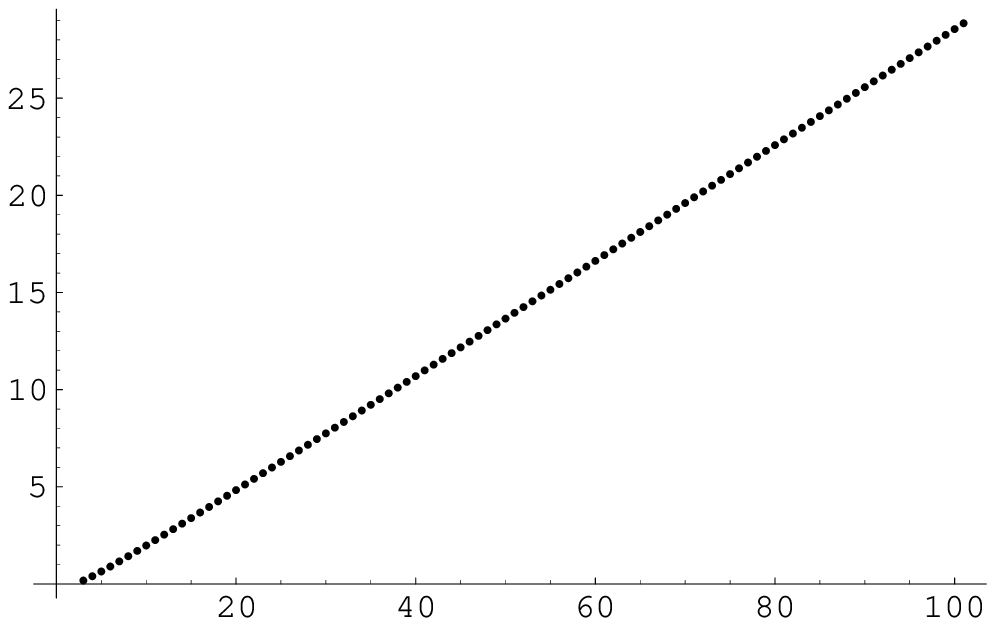}
\end{picture}
\caption{\label{figur1c}$g^{(1)}(v)$}
\end{minipage}
\begin{minipage}[]{5.0cm}
\vspace{5.0cm}
\begin{picture}(5.0,5.0)
\includegraphics[height=5.0cm,width=5.0cm]{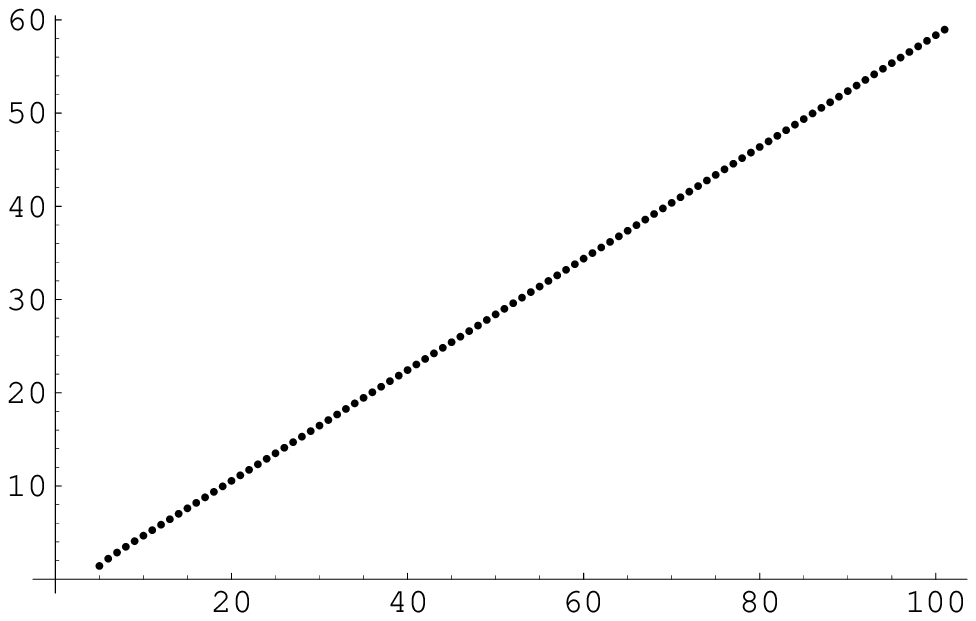}
\end{picture}
\caption{\label{figur1d}$g^{(2)}(v)$}
\end{minipage}
\begin{minipage}[h]{5.0cm}
\vspace{5.0cm}
\begin{picture}(5.0,5.0)
\includegraphics[height=5.0cm,width=5.0cm]{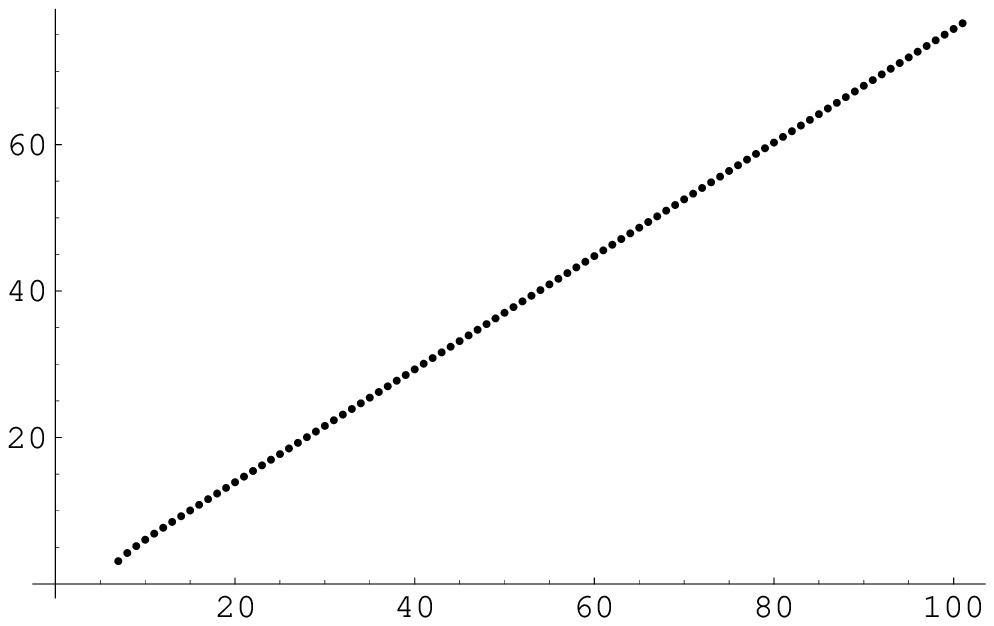}
\end{picture}
\caption{\label{figur1e}$g^{(3)}(v)$}
\end{minipage}
\begin{minipage}[]{5.0cm}
\vspace{5.0cm}
\begin{picture}(5.0,5.0)
\includegraphics[height=5.0cm,width=5.0cm]{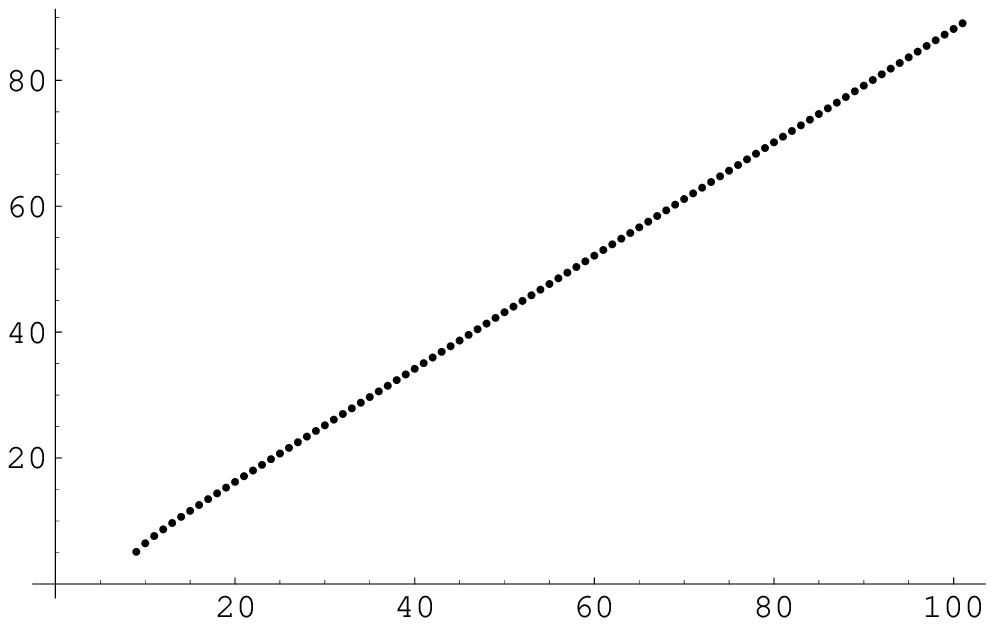}
\end{picture}
\caption{\label{figur1f}$g^{(4)}(v)$}
\end{minipage}
\begin{minipage}[]{5.0cm}
\vspace{5.0cm}
\begin{picture}(5.0,5.0)
\includegraphics[height=5.0cm,width=5.0cm]{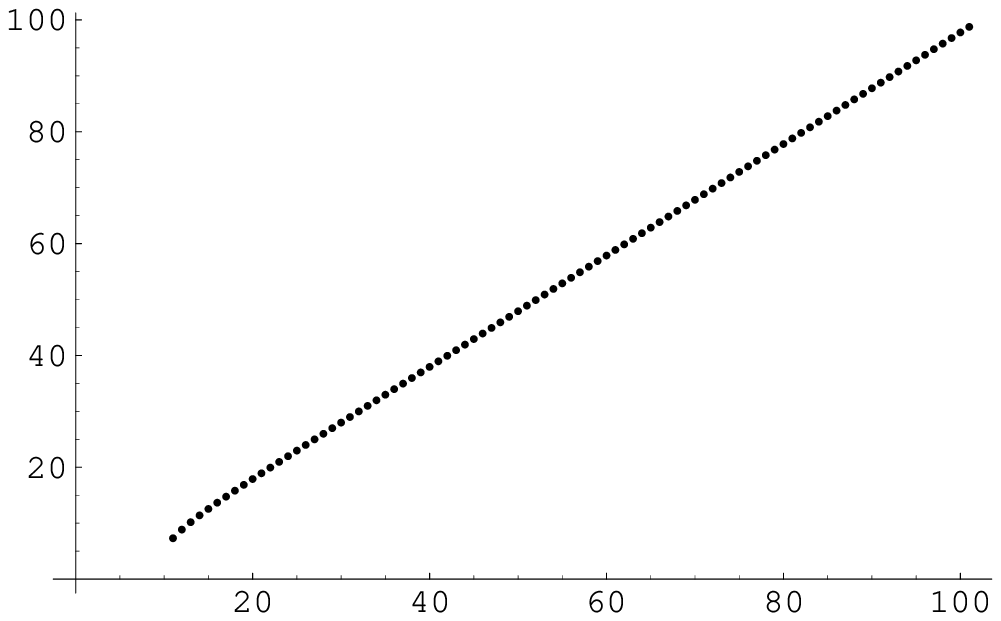}
\end{picture}
\caption{\label{figur1g}$g^{(5)}(v)$}
\end{minipage}
\begin{minipage}[]{5.0cm}
\vspace{5.0cm}
\begin{picture}(5.0,5.0)
\includegraphics[height=5.0cm,width=5.0cm]{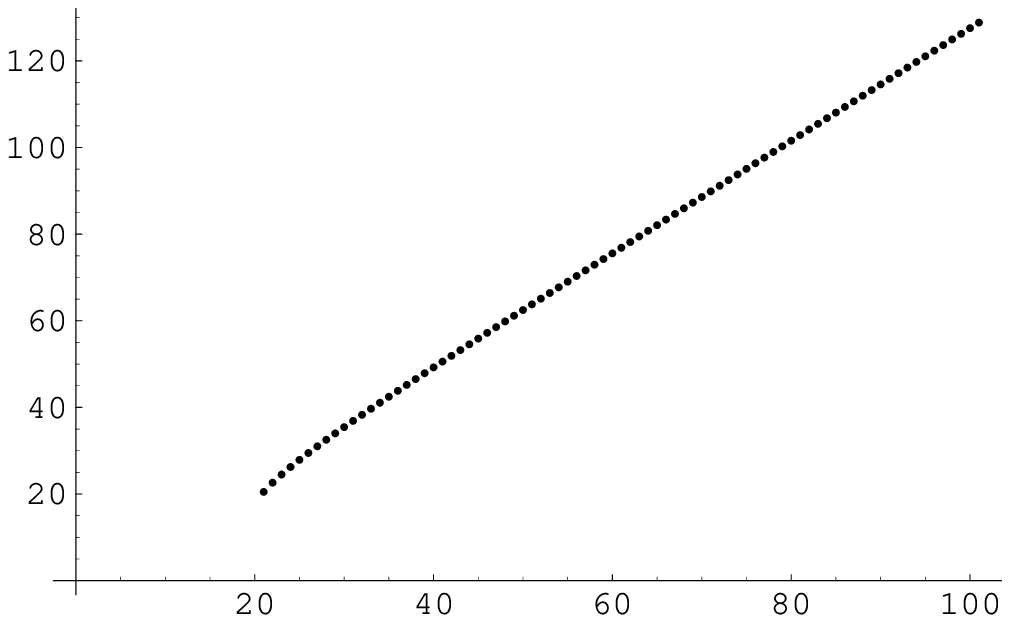}
\end{picture}
\caption{\label{figur1h}$g^{(10)}(v)$}
\end{minipage}
\end{center}
\end{figure}

\begin{figure}
\begin{center}
\begin{minipage}[h]{5.0cm}
\begin{picture}(5.0,5.0)
\includegraphics[height=5.0cm,width=5.0cm]{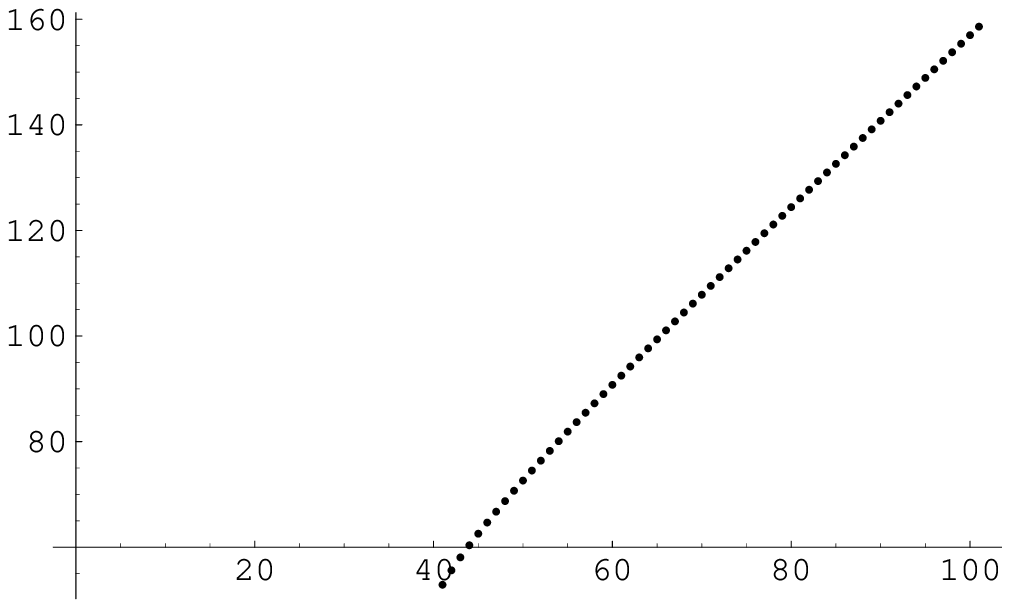}
\end{picture}
\caption{\label{figur1x}$g^{(20)}(v)$}
\end{minipage}
\vspace{1.5cm}
\begin{minipage}[]{5.0cm}
\begin{picture}(5.0,5.0)
\includegraphics[height=5.0cm,width=5.0cm]{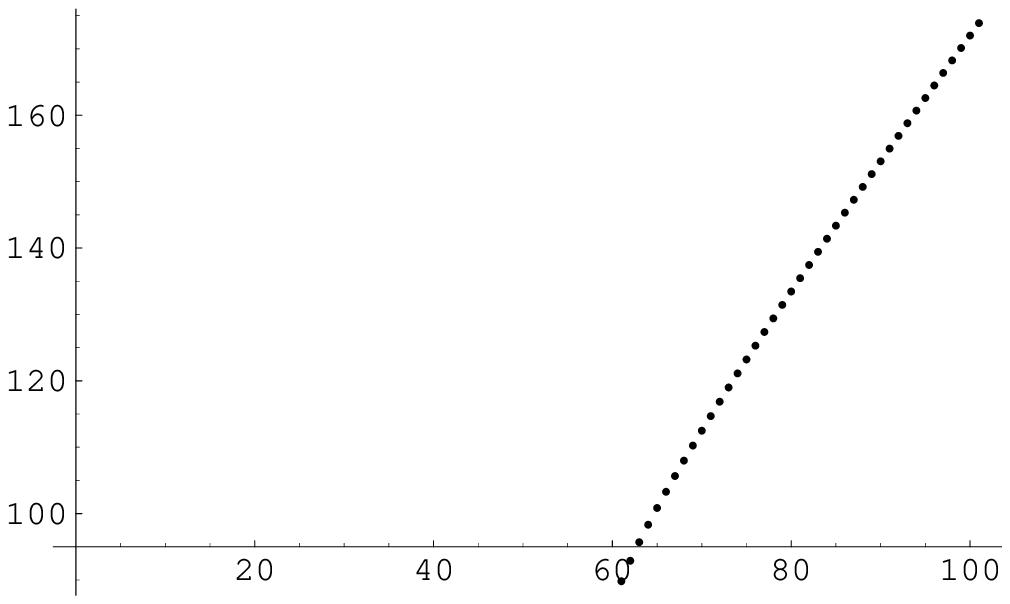}
\end{picture}
\caption{\label{figur1y}$g^{(30)}(v)$}
\end{minipage}
\begin{minipage}[]{5.0cm}
\vspace{5.0cm}
\begin{picture}(5.0,5.0)
\includegraphics[height=5.0cm,width=5.0cm]{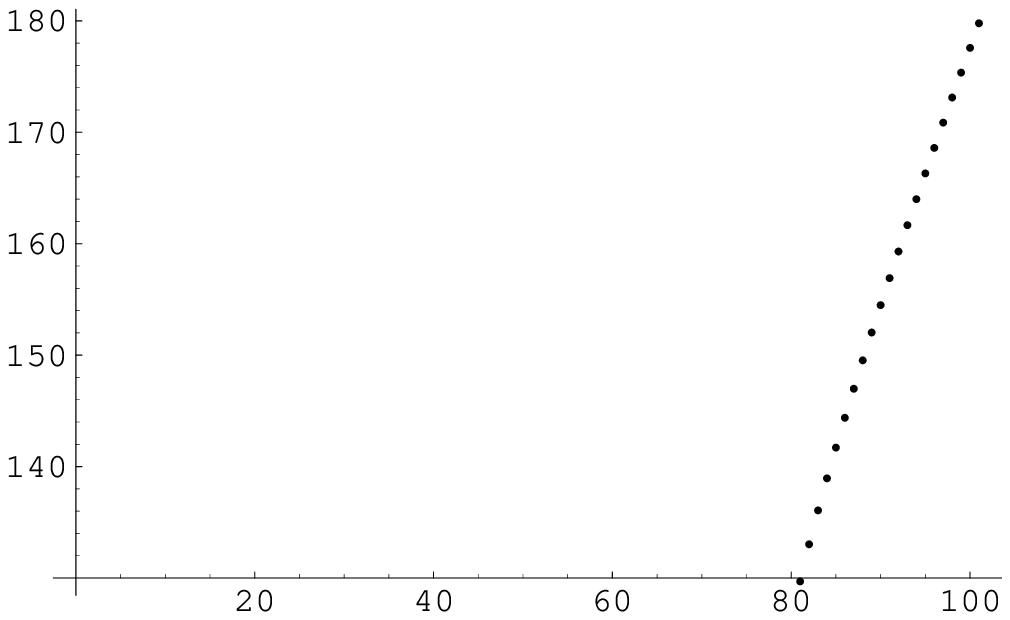}
\end{picture}
\caption{\label{figur1z}$g^{(40)}(v)$}
\end{minipage}
\vspace{1.5cm}
\begin{minipage}[]{5.0cm}
\vspace{5.0cm}
\begin{picture}(5.0,5.0)
\includegraphics[height=5.0cm,width=5.0cm]{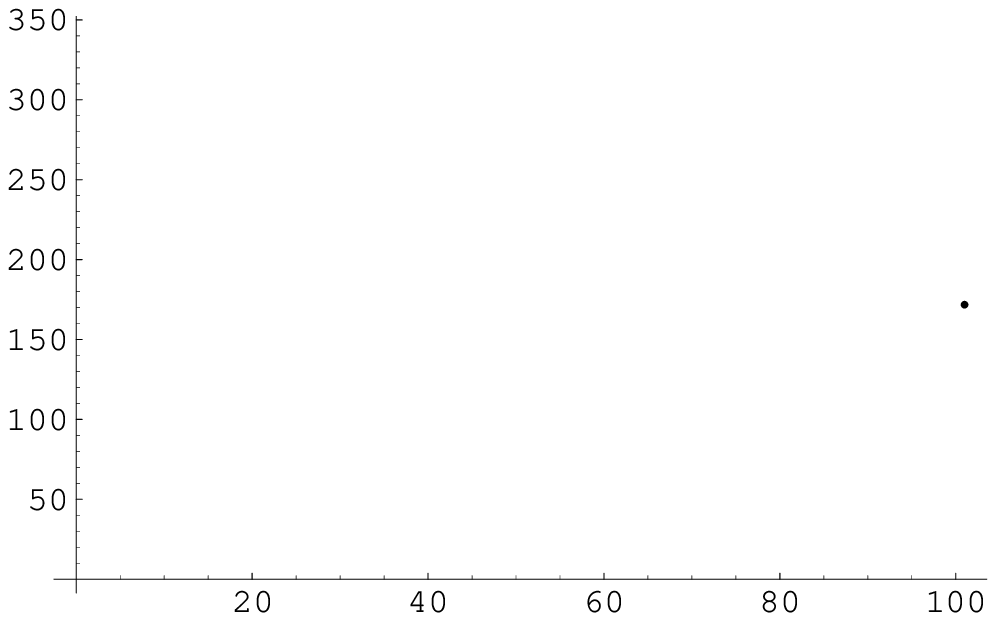}
\end{picture}
\caption{\label{figur1t}$g^{(50)}(v)$}
\end{minipage}
\end{center}
\end{figure}



\begin{thebibliography}{99}

\bibitem{sa} T. Richardson, R. Urbanke, {\textrm{http://lthcwww.epfl.ch/papers/ics.ps}}
\bibitem{da} R. Segdewick, P. Flajolet, ``An Introduction to The Analysis of Algorithms", Addison-Wesley Publishing Company, Inc., 1996
\bibitem{ea} R. Graham, D. Knuth, O. Patashnik, ``Concrete Mathematics", Addison-Wesley Publishing Company, Inc., 1994
\bibitem{fa} G. H. Hardy, ``A Course of Pure Mathematics", Cambridge University Press, Reprinted 1999
\bibitem{ga} E. C. Titchmarsh, ``The Theory of Functions", Oxford University Press, 1939
\bibitem{ba} E. Zauderer, ``Partial Differential Equations of Applied Mathematics", John Wiley \& Sons, Inc., 1998
\bibitem{ka} D. Betounes, ``Differential Equations: Theory and Applications", Springer-Verlag New York, Inc., 2001
\bibitem{xa} F. W. J. Olver, ``Asymptotic and Special Functions" Academic Press, 1974
\bibitem{wa} T. M. Apostol, ``Mathematical Analysis", Addison-Wesley Publishing Company, Inc., 1974

\end{thebibliography}
\end{document}